\documentclass[seceq,preprint]{ptptex}

\usepackage{graphicx}

\usepackage{amscd}

\def\a{\alpha}

\def\th{\theta}

\def\c{\raise2pt\hbox{$\chi$}}

\def\sfrac#1#2{{\textstyle\frac{#1}{#2}}}
\def\+{\dagger}
\def\={\ =\ }
\def\pa{\partial}

\def\>{\rangle}
\def\<{\langle}

\newcommand{\unity}{\boldsymbol{1}}

\newcommand{\e}{\mathrm{e}}
\newcommand{\im}{\,\mathrm{i}\,}
\newcommand{\diff}{\mathrm{d}}
\newcommand{\tr}{\mathrm{tr}}

\newcommand{\SU}{\mathrm{SU}}
\newcommand{\U}{\mathrm{U}}

\newcommand{\R}{{\mathbb{R}}}
\newcommand{\C}{{\mathbb{C}}}

\newcommand{\cA}{{\mathcal A}}
\newcommand{\cD}{{\mathcal D}}
\newcommand{\cE}{{\mathcal E}}
\newcommand{\cF}{{\mathcal F}}
\newcommand{\cH}{{\mathcal H}}
\newcommand{\cL}{{\mathcal L}}
\newcommand{\cM}{{\mathcal M}}

\newcommand{\be}{\begin{equation}}
\newcommand{\ee}{\end{equation}}
\newcommand{\bea}{\begin{eqnarray}}
\newcommand{\eea}{\end{eqnarray}}
\newcommand{\bal}{\begin{aligned}}
\newcommand{\eal}{\end{aligned}}

\newcommand{\und}{\qquad\text{and}\qquad}

\newcommand{\zb}{{\bar{z}}}
\newcommand{\ab}{{\bar{a}}}
\newcommand{\bb}{{\bar{b}}}

\preprintnumber[3cm]{ITP-UH-12/07\\[-6pt] HWM-07-13\\[-6pt] EMPG-07-09} 
\title{%        %You can use \\ for explicit line-break
Quiver Gauge Theory and Noncommutative Vortices%
}

\author{%       %Use \scshape  for the family name
Olaf \textsc{Lechtenfeld}${}^1$\ ,\quad
Alexander D.\ \textsc{Popov}${}^{1,2}$ \ and \ 
Richard J.\ \textsc{Szabo}${}^3$
}

\inst{%     %Affiliation, neglected when [addenda] or [errata]
\noindent ${}^1$
Institut f\"ur Theoretische Physik, Leibniz Universit\"at Hannover\\
Appelstra\ss{}e 2, 30167 Hannover, Germany
\\[3mm]
\noindent ${}^2$
Bogoliubov Laboratory of Theoretical Physics, JINR\\
141980 Dubna, Moscow Region, Russia
\\[3mm]
\noindent ${}^3$
Dept.\ of Mathematics and Maxwell Institute
for Mathematical Sciences\\ Heriot-Watt University,
Riccarton, Edinburgh EH14 4AS, U.K.
}

\abst{%         %this abstract is neglected when [addenda] or [errata]
\noindent
We construct explicit BPS and non-BPS solutions of the Yang-Mills
equations on noncommutative spaces $\R^{2n}_\theta\times G/H$ which are
manifestly $G$-symmetric. Given a $G$-representation, by twisting with
a particular bundle over $G/H$, we obtain a $G$-equivariant U($k$) bundle
with a $G$-equivariant connection over $\R^{2n}_\theta\times G/H$. The U($k$)
Donaldson-Uhlenbeck-Yau equations on these spaces reduce to vortex-type
equations in a particular quiver gauge theory on $\R^{2n}_\theta$.
Seiberg-Witten monopole equations are particular examples.
The noncommutative BPS configurations are formulated with partial
isometries, which are obtained from an equivariant Atiyah-Bott-Shapiro
construction. They can be interpreted as D0-branes inside a
space-filling brane-antibrane system.
\vskip1cm \small\noindent
Talk by O.L. at the 21st Nishinomiya-Yukawa Memorial Symposium, 
Kyoto, 15 Nov.\ 2006
}

\begin{document}

\maketitle

\section{Twisted dimensional reduction}

It is an old dream to 
``explain'' the standard model of particle physics by dimensional reduction
of a higher-dimensional gauge theory. After the reduction, the
field dependence on the extra coordinates must of course disappear from the 
four-dimensional Lagrangian. Usually, this is achieved, in a rather crude way,
by simply discarding the fields' dependence on the extra coordinates.
However, independence is by no means necessary: it suffices to prescribe 
{\it some\/} dependence, like, e.g., in warped compactifications.
If the extra spacetime dimensions admit isometries, it is particularly
elegant to compensate these by gauge transformations. In this way, the
Lie derivative with respect to a Killing vector becomes a gauge generator.
The bonus is a unification of gauge and Higgs sectors in the higher-dimensional
gauge theory.

The natural setting for spacetime isometries are coset spaces $G/H$,
and thus one is led to a reduction
$\cM\times\frac GH\ \longrightarrow\ \cM$
%\be
%\cM\times\frac GH\quad\longrightarrow\quad\cM
%\ee
where the manifold $\cM$ is to be specified later.
Such a ``coset-space dimensional reduction''~\cite{Kapetanakis:1992hf} 
was first suggested by Witten,\cite{Witten:1976ck} 
Forgacs and Manton,\cite{Forgacs:1979zs,Manton:1981es} and has since been 
extended supersymmetrically\cite{Manousselis:2000aj} and 
embedded into superstring theory.\cite{LopesCardoso:2002hd} \
In the present talk, for Lie groups $G$ of rank one and rank two,
we shall apply this scheme to perform a $G$-equivariant
reduction of Yang-Mills theory over $G/H$ to a quiver 
gauge theory on~$\cM$,\cite{Garcia-Prada:1993qv,Alvarez-Consul:2001mb,Alvarez-Consul:2001uk,Alvarez-Consul:2001um} 
formulate its BPS equations and show how to construct a certain class of
solutions, which admit a D-brane interpretation. These solutions, however, 
only exist when the system is subjected to a noncommutative deformation. 
Therefore, about half-way into the talk we specialize to $\cM=\C^n$ and 
apply a Moyal deformation. Most material presented here has appeared in 
Refs.~\citen{Lechtenfeld:2003cq,Popov:2005ik,Lechtenfeld:2006wu},
some is work in progress.

\section{K\"ahler times coset space $G/H$}

To be concrete, let us consider 
U($k$) Yang-Mills theory on $\cM_{2n}\times\frac GH$,
with $\cM_{2n}$ being a real $2n$-dimensional K\"ahler manifold with
K\"ahler form \ $\omega$ \ and metric \ $g$.
For cosets, we shall examine the following four examples:\\[4pt]
\begin{tabular}{lcccc} 
$G/H$:\quad{} 
& $\C P^1$ & $\C P^1\times\C P^1$ & $\C P^2$ & $Q_3$ \\
& $\|$ & $\|$ & $\|$ & $\|$ \\[4pt]
& \quad $\frac{\SU(2)}{\U(1)}$ \quad{} &
  \quad $\frac{\SU(2)\times\SU(2)}{\U(1)\times\U(1)}$ \quad &
  \quad $\frac{\SU(3)}{\mathrm{S}(\U(2)\times\U(1))}$ \quad &
  \quad $\frac{\SU(3)}{\U(1)\times\U(1)}$ \quad \\[8pt]
& $d{=}2$ & $d{=}4$ & $d{=}4$ & $d{=}6$
\end{tabular}
\\[12pt]
These are homogeneous but not necessarily symmetric spaces ($Q_3$ is not). 
Furthermore, they are K\"ahler, with K\"ahler forms $\beta\wedge\bar\beta$
factorized into canonical one-forms.

\section{Donaldson-Uhlenbeck-Yau equations}

To formulate U($k$) Yang-Mills theory on $\cM_{2n}\times\frac GH$,
we introduce a rank-$k$ hermitian vector bundle 
\be
\begin{CD} \cE \\ @VV \C^k V \\ \cM_{2n}\times\frac GH \end{CD}
\ee
with structure group $\U(k)$ and a connection $\cA$ which gives rise to the
curvature or field strength \ $\cF \= \diff\cA + \cA\wedge\cA$ \ 
subject to the Bianchi identity \ $D_{\!\cA}\,\cF=0$ \ 
where $D_{\!\cA}$ is the gauge covariant derivative.

The (vacuum) Yang-Mills equations read
\be \label{YM}
D_{\!\cA}\,(*\cF) \= 0
\ee
where `$*$' denotes the Hodge dual.
With respect to the K\"ahler form $\Omega=\omega+\beta{\wedge}\bar\beta$
of the total space, the field strength decomposes as
\be
\cF \= \cF^{(2,0)}\ +\ \cF^{(1,1)}\ +\ \cF^{(0,2)}\ .
\ee
So-called stable bundles $\cE$ solve the Donaldson-Uhlenbeck-Yau equations
\cite{donaldson,uhlenbeckyau}
\be
\cF^{(2,0)} \= 0 \= \cF^{(0,2)} \und *\Omega\wedge\cF \= 0
\qquad\qquad \text{(DUY)}
\ee
which are first-order conditions on the connection~$\cA$.
Their importance derives from the fact that the $n^2{-}n{+}1$ DUY equations
imply the $2n$ full Yang-Mills equations~(\ref{YM}).
Hence, for obtaining classical solutions it suffices to solve the DUY equations
rather than the full second-order field equations 
(but it is by no means necessary).
As a special case, on $\cM_4$ ($n{=}2$) the 3 DUY equations reduce to the 
famous self-duality equations $F=*F$ which yield instantons and monopoles.

\section{$G$-equivariant bundle construction}

In order to implement the coset-space reduction, we must construct a
$G$-equivariant bundle over the coset space. A rank-$d$ vector bundle \ 
$\cL \buildrel\C^d\over\longrightarrow G/H$ \ with structure group~U($d$)
is $G$-equivariant if the left translations $L_g$ on~$\cL$ (with $g\in G$)
are compatible with the right U($d$) action and the following diagram is 
commutative,
\be
\begin{CD} \cL @> L_g >> \cL \\
           @VV \pi V @V \pi VV \\
           \frac GH @> l_g >> \frac GH
\end{CD} \quad ,
\ee
where $l_g$ is the left translation on the coset space. Since $L_h\in\U(d)$ 
for $h\in H$, this defines a representation $\rho: H\hookrightarrow\U(d)$.
For simplicity, we take $\rho$ to be irreducible.

As a next step, we extend the bundle~$\cL$ by a rank-$k$ vector bundle~$E$
over $\cM_{2n}$,
\be
\begin{CD}\cL \\ @VV \C^d V \\ \frac GH \end{CD}
\qquad\longrightarrow\quad
\begin{CD}E\otimes\cL\\ @VV\C^k\otimes\C^d V\\ \cM_{2n}\times\frac GH\end{CD}
\qquad,
\ee
to a bundle over the total space with a trivial $G$-action on~$E$.
Further, we form a Whitney sum of $m{+}1$ such bundles
with data $(k_i,d_i,\rho_i)$ for $i=0,1,\dots,m$.
The $G$-equivariant total bundle 
\be
\cE \= \bigoplus_{i=0}^m E_i \otimes \cL_i
\ee
comes with a structure group $\prod_i\U(k_i)\times\U(d_i)$ 
and admits $G$-equivariant connections $\cA$ 
(i.e.\ connections compatible with equivariance).

\section{$G$-equivariant connection}

Finally, we twist each subbundle \
$E_i\buildrel\C^{k_i}\over\longrightarrow\cM_{2n}$ \
%$\begin{CD} E_i \\ @VV \C^{k_i} V \\ \cM_{2n} \end{CD}$ 
with a connection \ $A^i \in u(k_i)$ \ by the homogeneous bundle \
$\cL_i\buildrel\C^{d_i}\over\longrightarrow G/H$ \ 
%$\begin{CD} \cL_i \\ @VV \C^{d_i} V \\ G/H \end{CD}$ 
with a connection $a^i$ in the Lie$H$-irrep~$\rho_i$.
Hence, the connection on $E_i\otimes\cL_i$ reads 
\be
\cA^i \= A^i\otimes\unity_{d_i}\ +\ \unity_{k_i}\otimes a^i 
\qquad\text{for}\quad i=0,1,\ldots,m \ .
\ee
It is important to realize that the $G$-action connects different $H$-irreps,
$\rho_i\buildrel g\over\to \rho_j$, so that the total connection
\be
\cA \= \bigoplus_{i=0}^m \cA^i\ +\ \text{``off-diagonal''}
\ee
is not block-diagonal.
$G$-equivariance then dictates the decomposition of the connection
into $k_id_i{\times}k_jd_j$ blocks as 
\be \label{blockA}
\cA \= \begin{pmatrix}
A^0{\otimes}\unity_{d_0} + \unity_{k_0}{\otimes}\,a^0 &
\phi_{01}\otimes\beta_{01} & \cdots & \phi_{0m}\otimes\beta_{0m} \\[16pt]
\phi_{10}\otimes\beta_{10} &
A^1{\otimes}\unity_{d_1} + \unity_{k_1}{\otimes}\,a^1 &
\cdots & \phi_{1m}\otimes\beta_{1m} \\[16pt]
\vdots & \vdots & \ddots & \vdots \\[16pt]
\phi_{m0}\otimes\beta_{m0} & \phi_{m1}\otimes\beta_{m1} & \cdots &
A^m{\otimes}\unity_{d_m} + \unity_{k_m}{\otimes}\,a^m
\end{pmatrix}
\ee
with size $k_i{\times}k_j$ Higgs fields
\be
\phi_{ij}\=\phi^\+_{ji}\ \in\mathrm{Hom}(\C^{k_j},\C^{k_i})
\ee
in the bi-fundamental representation of $\U(k_i)\times\U(k_j)$
and size $d_i{\times}d_j$ one-forms \ $\beta_{ij}=-\beta^\+_{ji}$ \
on $\frac GH$ built from the components of $\beta$ and $\bar\beta$.

This construction breaks the original gauge group
\be
\U({\textstyle\sum}_i k_id_i)\ \longrightarrow\ \prod_i \U(k_i)
\ee
via the Higgs effect. 
In the following we choose the collection \ $\{\rho_i\}$ \
to descend from some $G$-irrep $\cD$, i.e.
\be
\cD|_{_H} \= \bigoplus_{i=0}^m \rho_i\ .
\ee
It should be noted that the coset generators connect 
only particular pairs $(\rho_j,\rho_i)$ so that many one-forms $\beta_{ij}$
actually vanish.

\section{The quiver diagram}

The connection \
$\cA\ \sim\ \{ A^i,\phi_{ij} \}$ \ realizes a quiver gauge theory:
For each $H$-irrep $\rho_i$ draw one vertex, 
which carries a multiplicity space $\C^{k_i}$ and a connection $A^i\in u(k_i)$;
for each nonzero one-form $\beta_{ij}: \rho_j\to\rho_i$ draw an arrow from 
vertex~$j$ to vertex~$i$, 
which carries a Higgs field $\phi_{ij}: \C^{k_j}\to\C^{k_i}$. Abbreviating 
$\phi_{ij}\otimes\beta_{ij}=:\Phi_{ij}$ we obtain pictorially\footnote{
Our arrows point to the left for later agreement with the standard building
of weight diagrams from the highest weight downward.
This is opposite to the convention of 
\citen{Popov:2005ik} and \citen{Lechtenfeld:2006wu}
where instead $\Phi_{ji}$ was used.}
\begin{equation}
\cdots\ \buildrel i\over\bullet\ \ \buildrel\Phi_{ij}\over\longleftarrow\ \
\buildrel j\over\bullet\ \cdots
\end{equation}
as the building block for a quiver diagram. The most general such diagram may 
in fact be obtained from the above construction by deleting some of the 
vertices (and connecting arrows).\cite{Alvarez-Consul:2001um}
In the following, we discuss examples based on $G$ of rank one and rank two.

\section{Rank-one example}

We come to the basic example of
\be
\frac GH\ \cong\ \frac{\SU(2)}{\U(1)}\ \cong\ S^2_R
\qquad\text{with}\qquad \beta \= \frac{2R^2\diff y}{R^2+y\bar y}\ ,
\ee
where $R$ is the radius of the two-sphere and $y$ denotes its (complex)
stereographic coordinate.

The homogeneous bundle in question is the $q$-monopole bundle
\be
\cL^q\=\cL^{\otimes q} \qquad\text{with}\qquad
\begin{CD} \cL{=}S^3 \\ @VV S^1 V \\ S^2_R \end{CD}
\ee
and transition functions\ $(\frac{y}{\bar y})^{\frac q2}=\e^{\im q\varphi}$.
The $q$-monopole connection and field strength read
\be
a_q\=\frac q2\,\frac{\bar y\diff y-y\diff\bar y}{R^2+y\bar y}
\qquad\longrightarrow\qquad f_q\=-\frac{q}{4R^2}\,\beta\wedge\bar\beta
\qquad\text{with}\quad c_1\!=\mathrm{deg}\,\cL^q=q \ .
\ee

Let us pick an $\SU(2)$-irrep \ $\cD=\underbar{$m{+}1$}$ \ 
(i.e.~spin $\frac m2$)
so that the $\U(1)$ irreps are characterized by charges $q_i=m{-}2i$ for
$i=0,1,\ldots,m$, and $\beta_{ij}=0$ except for $\beta_{i\ i-1}=-\beta$
and $\beta_{i\ i+1}=\bar\beta$. 
Labelling the vertices from the highest $\SU(2)$ weight downwards, 
we get chains
\be
\!\!E_m \buildrel\phi_{m\,m-1}\over\longleftarrow\!\!
\cdots \buildrel\phi_{21}\over\longleftarrow 
E_1 \buildrel\phi_{10}\over\longleftarrow E_0
\qquad\text{and}\qquad 
\cL^{-m} \buildrel\beta\over\longleftarrow
\cdots   \buildrel\beta\over\longleftarrow
\cL^{m-2}\buildrel\beta\over\longleftarrow \cL^m
\ee
which are represented diagrammatically by the linear (or $A_{m+1}$) quiver
\begin{equation}
\begin{CD} \buildrel m\over\bullet @< \ \!\Phi_{m\,m-1}\! <<
           \cdots @< \ \Phi_{32}\ <<
           \buildrel 2\over\bullet @< \ \Phi_{21}\ <<
           \buildrel 1\over\bullet @< \ \Phi_{10}\ <<
           \buildrel 0\over\bullet \end{CD} \quad.
\end{equation}

\section{Rank-two examples}

More instructive are the three rank-two examples listed in \S2.
\begin{tabular}{l} 
\hspace{-9mm}\normalsize First, in the product case of 
\\[10pt] \normalsize
$\C P^1\times\C P^1\ \cong\ \frac{\SU(2)}{\U(1)}\times\frac{\SU(2)}{\U(1)}$
\\[10pt] 
\hspace{-9mm}\normalsize the $G$-irrep is given by a pair of spins, 
\\[10pt] \normalsize 
$\cD\=(\underbar{$m_1{+}1$}\,,\,\underbar{$m_2{+}1$})$\ .
\\[10pt] 
\hspace{-9mm}\normalsize It is obvious that the corresponding quiver \\[2pt]
\hspace{-9mm}\normalsize becomes a product of two chains (see right).\\[2pt]
\hspace{-9mm}\normalsize Second, for the nonsymmetric coset 
\end{tabular}
\hfill
$\begin{CD} \bullet @<<< \bullet @<<< \bullet @<<< \cdots @<<< \bullet \\
            @VVV @VVV @VVV & & @VVV \\
            \bullet @<<< \bullet @<<< \bullet @<<< \cdots @<<< \bullet \\
            @VVV @VVV @VVV & & @VVV \\
            \vdots & & \vdots & & \vdots & & & & \vdots \\
            @VVV @VVV @VVV & & @VVV \\
            \bullet @<<< \bullet @<<< \bullet @<<< \cdots @<<< \bullet
\end{CD}$
%Second, for the nonsymmetric coset
\be
Q_3\ \cong\ \frac{\SU(3)}{\U(1)\times\U(1)}
\ \cong\ \frac{G}{\mathrm{maximal~torus}}
\ee
the $\rho_i$ are labelled by the eigenvalues of the $\SU(3)$ Cartan generators,
and thus the quiver is simply based on the {\it weight\/} diagram of the 
$\SU(3)$ representation $\cD$.
We order the weights descending from the highest one, and our arrows
agree with the action of the lowering operators. Third, the case of
\be
\C P^2\ \cong\ \frac{\SU(3)}{\mathrm{S}(\U(2)\times\U(1))}
\ee
calls for a decomposition \ $\cD = \bigoplus_i \underbar{$d_i$}_{\ q_i}$ \
of the $\SU(3)$ representation into `isospin' irreps $\underbar{$d_i$}$ with
`hypercharge'~$q_i$ and a $(d_i,q_i)$ plot for the quiver vertices.
Since each vertex represents a full isospin multiplet, we may alternatively
obtain the corresponding quiver diagram for $\C P^2$ from the $Q_3$ quiver
by collapsing all vertices of a `horizontal' $\SU(2)$-irrep to single vertex.

Clearly, the novel features of the rank-two situation are, firstly,
the appearance of multiple arrows due to weight degeneracy and, secondly,
the occurrence of nontrivial Higgs-field relations, 
such as $\Phi_{32}\Phi_{21}=\Phi_{31}$,
due to the commutativity of the quiver diagrams.

\begin{figure}
\centerline{\includegraphics[width=14 cm]{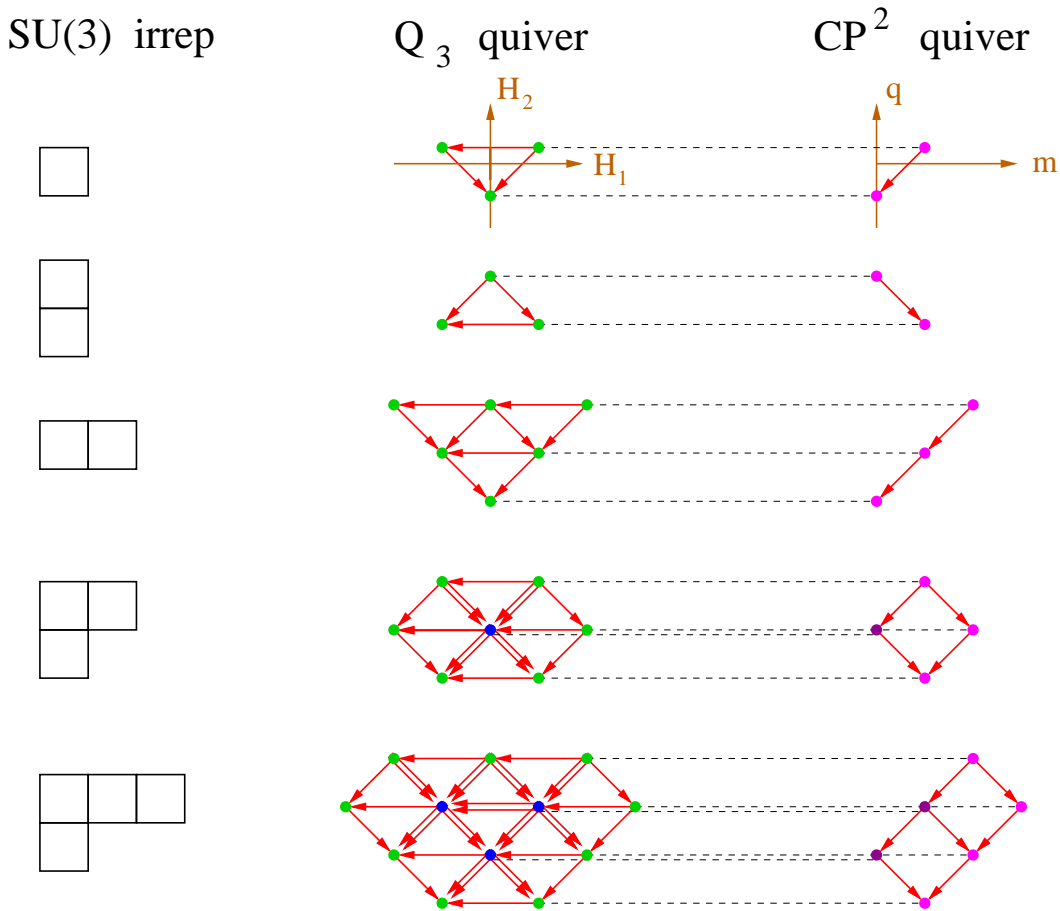}}
\caption{SU(3) irreps, weights and quiver diagrams}
\label{fig:1}
\end{figure}

\section{Nonabelian coupled vortex equations}

The condition of $G$-equivariance together with the data $\{\cD, k_i\}$
uniquely determine the dependence of $\cA$ and $\cF$ on the coset coordinates.
Therefore, the Yang-Mills and DUY equations dimensionally reduce to equations
for $A^i$ (or $F^i$) $\in u(k_i)$ 
and $\phi_{ij}\in\text{Hom}(\C^{k_j},\C^{k_i})$ on $\cM_{2n}$ only,
with the indices $i,j=0,1,\ldots,m$ running over the vertices of the quiver and
index pairs $(i,j)$ labelling the blocks in~(\ref{blockA}).
For explicitness, we introduce local holomorphic coordinates
$\{z^a\}$ with $a=1,2,\dots,n$ on $\cM_{2n}$,
so that the $\U(k_i)$ connection and field strength take the form
\be
A^i \= A^i_a\,\diff z^a + A^i_\ab\,\diff \zb^\ab
\qquad\longrightarrow\qquad
F^i \= F^i_{ab}\,\diff z^a{\wedge}\diff z^b +
     2 F^i_{a\bb}\,\diff z^a{\wedge}\diff\zb^\bb +
       F^i_{\ab\bb}\,\diff\zb^\ab{\wedge}\diff\zb^\bb
\ee       
with \ $(A^i_a)^\+=-A^i_\ab$ \ and \ $(F^i_{ab})^\+=F^i_{\ab\bb}$, \
$(F^i_{a\bb})^\+=F^i_{\ab b}$ \ etc..
For the rank-one case with \ $\cD=\underbar{$m{+}1$}$ \
and redenoting \ $\phi_{ij}=:\phi_{i,j}$, the DUY equations
on \ $\cM_{2n}\times\C P^1$ \ descend to
\bea \label{chain1}
& F^i_{ab} \= 0 \= F^i_{\ab\bb} \quad,\qquad
D_\ab\,\phi_{i,i+1} \= 0 \= D_a\,\phi_{i+1,i} \quad, \\[8pt] \label{chain2}
& g^{a\bb}\,F^i_{a\bb} \= \sfrac{1}{2R^2}
\bigl({m{-}2i}\ +\ \phi_{i,i-1}\phi_{i-1,i}\ -\ \phi_{i,i+1}\phi_{i+1,i}
\bigr) \ ,
\eea
where $D$ denotes the gauge covariant derivative, and $\phi_0=\phi_{m+1}=0$.
We call this set of relations the ``nonabelian chain vortex equations'' 
with data\ $(\cM_{2n},R,m,\{k_i\})$.

\section{Seiberg-Witten monopole equations}

The simplest nontrivial case occurs for \ 
$\cM_4$ (i.e.~$n{=}2$), a spin-$\frac12$ representation (i.e.~$m{=}1$) 
and the breaking $\U(2)\to\U(1)\times\U(1)$.
Dropping irrelevant indices, $\unity$s and $\otimes$s, the connection becomes
\be
\cA \= \begin{pmatrix}
A^0(z)+a_{+1}(y) & \phi(z)\,\bar\beta(y) \\[24pt]
-\bar\phi(z)\,\beta(y) & A^1(z)+a_{-1}(y)
\end{pmatrix} \ .
\ee
The DUY equations then imply \ $A^0=-A^1=:A$ \ and simplify to
\be
F_{ab} \= 0 \= F_{\ab\bb} \quad,\qquad
\pa_\ab\,\phi + 2 A_\ab\,\phi \= 0 \quad,\qquad
g^{a\bb}\,F_{a\bb} \= \sfrac{1}{2R^2} \bigl(1-\phi\,\bar\phi\bigr) \ ,
\ee
which are known as the 
``perturbed abelian Seiberg-Witten monopole equations''.\cite{Witten:1994cg}
On $\cM_4=\R^4$, the latter admit only trivial solutions; one of the reasons 
why we shall now apply a noncommutative deformation.\cite{Popov:2003xg}

\section{Moyal deformation}

For the remainder of the talk we specialize to \ $\cM_{2n}=\C^n$ \ 
in order to Moyal deform the base manifold. This deformation 
is realized by the Moyal-Weyl map sending
\bea
\text{Schwartz functions $f$} & \qquad\longmapsto\qquad & 
\text{compact operators $\widehat{f}$} \\
\text{coordinates $z^a$ and $\bar z^\bb$} & \qquad\longmapsto\qquad & 
\text{operators $\widehat{z}^a$ and $\widehat{\bar z}\vphantom{z}^\bb$} 
\eea
subject to \ $[\widehat{z}^a,\widehat{\bar z}\vphantom{z}^\bb] = \th^{a\bb}$ \
with an antisymmetric matrix~$(\th^{a\bb})$.
We can always rotate the coordinates such that 
\be
\th^{a\bb} \= 2\delta^{ab}\,\th^a \qquad\text{for}\quad
\th^a\in\R_+ \quad\text{with}\quad a,b=1,\dots,n\ .
\ee
This defines the noncommutative space \ $\C^n_\th$, 
with isometry $\mathrm{USp}(n)$ and 
carrying $n$ copies of the Heisenberg algebra,
\be
\bigl[ \sfrac{\widehat{z}^a}{\sqrt{2\th^a}}\,,\,
\sfrac{\widehat{\bar z}\vphantom{z}^\bb}{\sqrt{2\smash{\th^b}}} \bigr]
\= \delta^{ab} \ .
\ee
To represent this algebra, we need to introduce an auxiliary Fock space~$\cH$.
Finally, we remark that derivatives and integrals are represented as follows
($\th^{a\bb}\,\th_{\bb c}=\delta^a_c$),
\be
\pa_\bb f \ \longmapsto \
\th_{\bb c}\,[\widehat{z}^c,\widehat{f}] 
\qquad\text{and}\qquad
\smallint\!\diff\mathrm{V}\,f \ \longmapsto \
(2\pi)^n\mathrm{Pf}(\th)\,\tr_\cH \widehat{f}\ .
\ee

\section{Noncommutative chain vortex system}

How do the nonabelian chain vortex equations (\ref{chain1}, \ref{chain2}) 
change under the Moyal deformation? Dropping the hats from now on, we
define ``covariant coordinates''
\be
X^i_a \ :=\ A^i_a\ +\ \th_{a\bb}\,\zb^\bb \qquad\text{and}\qquad
X^i_\ab\ :=\ A^i_\ab\ +\ \th_{\ab b}\,z^b 
\ee
and express the field strengths and Higgs gradients through them,
\be
F^i_{ab} \= [X^i_a,X^i_b] \ , \quad
F^i_{a\bb} \= [X^i_a,X^i_\bb]\ +\ \th_{a\bb} \quad\text{and}\quad
D_\ab\,\phi_{i,i+1}\=X^i_\ab\,\phi_{i,i+1}-\phi_{i,i+1}X^{i+1}_\ab\ .
\ee
With this, the DUY/vortex equations (\ref{chain1}, \ref{chain2})
reduce to algebraic equations for $\{X^i,\phi_{i,i+1}\}$:
\bea \label{nchain1}
& [X^i_a,X^i_b] \= 0 \= [X^i_\ab,X^i_\bb] \quad,\qquad
X^i_\ab\,\phi_{i,i+1}\ -\ \phi_{i,i+1}X^{i+1}_\ab \= 0 \quad,\\[8pt] 
& \delta^{ab}\,\bigl( [X^i_a,X^i_\bb]+\th_{a\bb} \bigr) \= \sfrac{1}{4R^2}
\bigl(m{-}2i\ +\ \phi_{i,i-1}\phi_{i-1,i}\ -\ \phi_{i,i+1}\phi_{i+1,i}
\bigr)\ . \label{nchain2}
\eea

\section{BPS solutions}

We remain with the $\frac GH = \C P^1$ case and
consider momentarily the particular situation of $k_1=\ldots=k_m=:r$,
i.e.\ gauge group \ $\U(k_0)\times\U(r)^m$.
In this context, a good ansatz is
\bea \label{ans1}
&& A^i_a \= 0 \und \phi_{i,i+1} \sim\ \unity_r
\qquad\text{for}\quad i=1,2,\dots,m  \\[8pt] \text{but}\qquad
&& A^0_a \=\th_{a\bb}\,\bigl(T\,\zb^b\,T^\+\ -\ \zb^b\bigr)
\und \phi_{0,1} \= \sqrt{m}\;T \ , \label{ans2}
\eea
with a partial isometry realized by a $k_0{\times}r$ matrix $T$
(Toeplitz operator) obeying 
\be
T^\+T\=\unity_r \ ,\qquad T\,T^\+\=\unity_{k_0}-P \ ,
\qquad P^2=P=P^\+ \quad\text{with}\quad \text{rk}(P)=:N \ .
\ee
Suitable operators $T$ obtain from an $\SU(2)$-equivariant generalization 
of the ABS construction~\cite{ABS}.
With this ansatz, the field strengths and Higgs gradients become
\be \label{BPSsol}
F^i_{\cdot\cdot}=0 \qquad\text{except}\qquad F^0_{a\bb}=\th_{a\bb}\,P
\qquad\text{and}\qquad D_\ab\,\phi_{i,i+1} =0= D_a\,\phi_{i,i+1}\ .
\ee
Finally, plugging the ansatz into the noncommutative chain vortex system
(\ref{nchain1}, \ref{nchain2}), we observe that all equations are fulfilled
provided 
\be \label{cond}
\sum_{a=1}^n \frac{1}{\th^a} \= \frac{m}{2R^2} \ ,
\ee
a nontrivial relation between the deformation strength and the size of the
coset space!

\section{Non-BPS solutions}

Turning on more than one quiver vertex in the ansatz above fails to
produce a nontrivial solution to the noncommutative DUY/vortex equations.
Nevertheless, let us consider the general situation of $\prod_i\U(k_i)$
as the gauge group and generalize the ansatz (\ref{ans1}, \ref{ans2}) to
\be
A^i_a \= \th_{a\bb}\,\bigl(T_i^{\phantom{\+}}\zb^b\,T_i^\+ - \zb^b\bigr)
\und \phi_{i,i+1} \= \a_{i+1}\,T_i^{\phantom{\+}}\,T_{i+1}^\+
\quad\text{with}\quad \a_i\in\C \ ,
\ee
where $m{+}1$ partial isometries are realized by $k_i{\times}r$ matrices
$T_i$ (Toeplitz operators): 
\be
T_i^\+ T_i^{\phantom{\+}}\=\unity_r \ ,\qquad
T_i^{\phantom{\+}}T_i^\+\=\unity_{k_i}-P_i^{\phantom{\+}} \ ,\qquad
P_i^2=P_i^{\phantom{\+}}=P_i^\+ \quad\text{of rank}\ N_i \ .
\ee
This ansatz implies
\bea
&& F^i_{ab}\=0\=F^i_{\ab\bb}\ ,\quad F^i_{a\bb}\=\th_{a\bb}\,P_i\ ,\quad
D_\ab\,\phi_{i,i+1} \=0\= D_a\,\phi_{i,i+1} \\[8pt]
&& \text{and}\qquad 
|\a_i|^{-2}\,\phi_{i,i-1}\,\phi_{i-1,i} \= \unity_{k_i}-P_i \=
|\a_{i+1}|^{-2}\,\phi_{i,i+1}\,\phi_{i+1,i}\ ,
\eea
which finally contradicts (\ref{nchain2}) 
if more than one projector is nonzero. 

Surprisingly, however, it
does solve the {\it full\/} noncommutative Yang-Mills equations!
The energy of the so-constructed non-BPS configurations is given by
\be
E \= 2\pi R^2 \bigl(\prod_{a=1}^n 2\pi\th^a\bigr) \sum_{i=0}^m\,\text{Tr}_\cH\,
\bigl[ \lambda_i\,P_i\ +\ \mu_i\,(\unity_{k_i}{-}P_i) \bigr]
\ee
\be
\text{with}\qquad \textstyle
\lambda_i \= \sum_b\frac{1}{(\th^b)^2}\ +\ \frac{(m{-}2i)^2}{4\,R^4} \und
\mu_i \= \frac{(m{-}2i{+}|\a_i|^2{-}|\a_{i+1}|^2)^2}{4\,R^4} \ ,
\ee
where $\a_0=\a_{m+1}=0$.
Finite energy requires $\mu_i=0$ for $i=0,1,\ldots,m$, which determines
$|\a_{i+1}|^2=(i{+}1)(m{-}1)$.
The BPS solution (\ref{BPSsol}) with (\ref{cond}) is seen as a special case:
Putting $P_1=\ldots=P_m=0$ (and $\mu_i=0$) yields
\be
E_{\text{BPS}} \= 
2\pi R^2 \bigl(\prod_{a=1}^n 2\pi\th^a\bigr)\,\lambda_0\, \text{Tr}_\cH\,P_0
\qquad\text{with}\qquad \lambda_0 \= 2\sum_{b\le c} (\th^b\th^c)^{-1}\ .
\ee

\section{D-brane interpretation}

Our construction and the constructed classical field configurations allow
for a D-brane interpretation. For simplicity, let us stay with the
$\frac GH = \C P^1$ case. One has a higher-dimensional and a lower-dimensional
picture:

``Upstairs'' on $\C^n_\th \times S^2$ we began with 
$k$~coincident D($2n{+}2$)-branes wrapping the $S^2$.
The $\SU(2)$-equivariance condition splits $k\to\{k_i\}$ 
and wraps the $S^2$ with charge-$q_i$ monopole fields,
for $i=0,\ldots,m$.

``Downstairs'' on $\C^n_\th$ we find
$m{+}1$ subsets of D($2n$)-branes carrying magnetic fluxes $q_i$.
On each subset of these space-filling branes live Chan-Paton gauge fields
$A^i\in\mathrm{End}(E_{k_i})$, and neighboring subsets are connected by
Higgs fields $\phi_{i,i+1}\in\mathrm{Hom}(E_{k_{i+1}},E_{k_i})$ which
correspond to massless open-string excitations.

This chain of brane subsets is marginally bound but stabilized by the
magnetic fluxes. The BPS vortex configurations we have constructed
are bound states of $mN$ D0-branes inside the D($2n$)-brane system.
The energy and topological charge of such a BPS state is most elegantly
computed via equivariant K-homology.

The aforesaid generalizes to quivers based on higher-rank Lie groups and 
their corresponding vortex-type equations, but some new features will arise 
due to nontrivial Higgs-field relations and quiver vertex degeneracies.

\section*{Acknowledgements}
O.L. thanks Lutz Habermann for clarifying the equivariant bundle construction.

\end{document}